\documentclass[showpacs, twocolumn]{revtex4-1}
\usepackage{graphicx}
\usepackage{amsmath}
\usepackage{appendix}
\begin{document}

\title{ Fluctuations and quantum self-bound droplets in a dipolar Bose-Bose mixture}

\author{Abdel\^{a}ali Boudjem\^{a}a$^{1,2}$}
\affiliation{$^1$Department of Physics, Faculty of Exact Sciences and Informatics, 
and $^2$Laboratory of Mechanics and Energy, Hassiba Benbouali University of Chlef, P.O.
Box 78, 02000, Ouled-Fares, Chlef, Algeria.}
\email {a.boudjemaa@univ-chlef.dz}

\date{\today}

\begin{abstract}

We systematically investigate the properties of three-dimensional dipolar binary Bose mixture at low temperatures.
A set of coupled self-consistent equations of motion are derived for the two condensates. 
In the homogeneous case, useful analytical formulas for the condensate depletion, the anomalous density, the ground-state energy,
and the equation of state are obtained.  
The theory is extended to the inhomogeneous case and the importance of the inhomogeneity is highlighted.
Our results open up a new avenue for studying dipolar mixture droplets. 
Impacts of the dipole-dipole interaction on the stability, density profiles, and the size of the self-bound droplet are deeply discussed.
The finite-temperature behavior of such a state is also examined.

\end{abstract}


\maketitle

\section{Introduction} \label{Intro}

Quantum degenerate gases of bosonic mixtures, achieved by using either different hyperfine states, different isotopes
of the same species or different atomic species have sparked a great interest from both theoretical and experimental studies.
These systems have proved to be an ideal platform for exploring quantum many-body physics 
due to their rigorous control of the inter- and intra-component interactions.

Bose-Bose mixtures with dipole-dipole interactions (DDIs) represents an interesting model 
for observing and understanding new states of matter in many areas of physics
due to their anisotropic and long-range interactions. 
Experimentally, binary dipolar Bose-Einstein condensates (BECs) can be created following different routes namely:
two different Rydberg states \cite{SSP} or heteronuclear diatomic molecules \cite {Eng, Debt, Pasq}.
Most recently, the first realization of a two-species magneto-optical trap for Er-Dy has been reported in \cite{Ilz}.

From the theoretical side, ground-state properties, the immiscibility-miscibility transition (IMT), and the phase separation of 
harmonically trapped two-component dipolar BECs have been investigated in \cite{Sto, Glig, Kui, Adhik}.
In quasi-two-dimensional (2D) geometry,  the IMT occurs due to the roton instability \cite{Wilson}. 
In a single component BEC, the roton instability may strongly enhance the quantum and the thermal fluctuations \cite{BoudjG}.
The competition between the inter- and intraspecies interactions leads to the emergence of nonlocal solitons in dual dipolar BECs \cite{Adhik1}.
It has been found that such systems may exhibit many interesting vortex structures, such as interlaced honeycomb
and octagonal vortex clusters, as well as vortex necklaces \cite {Shir, Xiao, Xue}.
Dipolar bosonic mixtures in optical lattices constitute also ideal candidates for the observation of the supersolid phase (see e.g. \cite{Wilson1, Yong}) 
owing to their large dipole moments and the high precision control over their internal and motional states.
Very recently, the properties of homogeneous 3D and 2D two-component BECs with DDIs  have been investigated using beyond mean-field theory
\cite{Past}.

Almost all previous works for binary dipolar mixtures have been limited to the case of zero temperature.
The aim of the present paper is to study the role of temperature effects in excitations, fluctuations, and thermodynamics of Bose mixtures with DDIs.
To this end, we employ the Hartree-Fock-Bogoliubov (HFB) theory. 
This scheme has been successfully utilized in 3D dipolar one-component systems with two- and three-body interactions \cite{Boudj2015, BoudjDp, Boudj2017}.
Our work reveals two important effects which are dissimilar to a single component BEC, (i) large condensate depletion 
(ii) near the phase separation and at low temperature, the thermal contribution to the depletion and all thermodynamic quantities has a distinct temperature dependence.  
This is most likely due to the intriguing interplay of inter- and intraspecies interactions. 
We show also that quantum and thermal fluctuations may significantly affect the excitations and the thermodynamics of the system even at very low temperatures.

On the other hand, recent theoretical and experimental studies of self-bound quantum droplets in nondipolar two-component BECs 
with competing attractive interspecies and repulsive intraspecies interactions \cite {Petrov, PetAst, Capp, Cab, Sem, Cik} 
open the possibility of entirely new prospects for ultracold atomic physics.
The formation of such an exotic state,  which survives even in free space, arises from a repulsive beyond mean-field Lee-Huang-Yang (LHY) term. 
Up to now, the effects of the DDI on the properties of a mixture self-bound droplet have remained unexplored.

Our motivation is then to investigate the formation of the self-bound droplet in a dipolar bosonic mixture of intraspecies repulsive interactions, 
and attractive interspecies interaction using our time-dependent-HFB (TDHFB) equations  \cite{Boudj, Boudj55, Boudj66, Boudj00}.
By precisely adjusting the strength of the DDI, we show that the repulsive LHY quantum corrections which provide an extra term, $\propto n_c^{5/2}$, 
arrest the attractive mean-field term, $\propto n_c^2$, enabling the nucleation of a stable dipolar mixture droplet.
This stabilization scenario resembles that which has occurred in nondipolar binary Bose-Bose mixtures \cite {Petrov, PetAst, Capp, Cab} and in
a dipolar one-component BEC \cite {BoudjDp,Pfau, Saito, Wach, Bess, Chom}. 
By sufficiently tuning the dipolar interaction below the $s$-wave scattering length, we find that the droplet becomes slightly anisotropic because of DDIs.
%but it is just a small perturbation to the  droplet in a nondipolar mixture with repulsive intra- and attractive interspecies interactions. 
We point out that the dipolar interactions lead to lowering the central density and the number of particles as well as they squeeze the droplet widths.
We then extend this study to the finite-temperature, by numerically solving our TDHFB equations. 
The condensate and noncondensate density profiles in the droplet are profoundly analyzed.
Our results show that the thermal fluctuations may modify the equilibrium of the droplet.

The paper is structured as follows. In Sec.\ref{Mod}, we introduce the basic formalism describing dipolar Bose mixtures.
We derive coupled equations of motion that enables us to study the dynamics of the two condensates using the Hartree-Fock-Bogoliubov (HFB) approximation.
In Sec.\ref{Flth}  we obtain useful formulas linking quantum and thermal fluctuations of some thermodynamic quantities, 
such as the chemical potential, the ground state energy and the compressibility for the homogeneous mixture.
In Sec.\ref{InhM} we generalize the theory to the case of the inhomogeneous Bose-condensed mixture with DDIs, 
using the local density approximation (LDA). 
Section \ref{Drop} deals with effects of the DDIs on the physics of the droplet state in dilute dipolar Bose mixtures at both zero and finite temperatures. 
Our results are summarized in Sec.\ref{concl}.

%This section recaps a few interesting findings reported in previous studies. 

\section{Model} \label{Mod}

We consider weakly interacting  two-component dipolar BECs with the atomic mass $m_j$.
The grand-canonical  Hamiltonian of the system reads as follows:
\begin{align}\label{ham}
\hat H& = \sum_{j=1}^2 \bigg[\int d {\bf r} \, \hat \psi_j^\dagger ({\bf r}) h_j^{sp} \hat\psi_j(\bf{r})  \\
&+\frac{1}{2} \int d {\bf r} \int d{\bf r'}\, \hat\psi_j^\dagger({\bf r}) \hat\psi_j^\dagger ({\bf r'}) V_j ({\bf r-r'})\hat\psi_j({\bf r'}) \hat\psi_j(\bf{r}) \bigg]\nonumber \\
&+\int d {\bf r} \int d{\bf r'}\, \hat\psi_1^\dagger({\bf r}) \hat\psi_2^\dagger ({\bf r'}) V_{12} ({\bf r-r'})\hat\psi_2 ({\bf r'}) \hat\psi_1(\bf{r}), \nonumber
\end{align}
where  $\hat\psi_j^\dagger$ and  $\hat\psi_j$ denote, respectively the usual creation and annihilation field operators, satisfying the usual canonical commutation rules 
$[\hat\psi_j({\bf r}), \hat\psi_j^\dagger (\bf r')]=\delta ({\bf r}-{\bf r'})$, and $h_j^{sp} =-(\hbar^2 /2m_j) \Delta +U_j({\bf r})-\mu_j$ is the single particle Hamiltonian, 
with $U_j({\bf r})$ being the external traps and $\mu_j$ representing chemical potentials related to each component.

The intraspecies two-body interaction potential  is 
\begin{equation}  \label{IntraPot}
V_j ({\bf r})=g_j\delta({\bf r})+d_j^2 \frac{1-3\cos^2\theta} { r^3},
\end{equation}
where $g_j=4\pi \hbar^2 a_j/m_j$ with $a_j$ being the intraspecies  $s$-wave scattering lengths.
The last term in Eq.(\ref{IntraPot}) accounts for the DDI potential where $d_j$ stands for the magnitude of the dipole moment of component $j$ and 
$\theta$ is the angle between the polarization axis and the relative separation of the two dipoles, it is supposed to be the same for both components.
The intraspecies dipole-dipole distance is defined as $r_{*j}= m_j d_j^2/\hbar^2$. \\
The interspecies two-body interactions potential reads
\begin{equation}  \label{InterPot}
V_{12}({\bf r})=g_{12}\delta({\bf r})+d_1 d_2 \frac{1-3\cos^2\theta} { r^3},
\end{equation}
where $g_{12}=g_{21}= 2\pi \hbar^2 (m_1^{-1}+m_2^{-1}) a_{12}$ corresponds to the interspecies 
short-range part of the interaction, which is characterized by the interspecies  $a_{12}=a_{21}$ $s$-wave scattering lengths.
The interspecies dipole-dipole distance is $r_{*12}=r_{*21}= 2d_1 d_2 /[\hbar^2  (m_1^{-1}+m_2^{-1})]$.

%Here the dipoles are supposed to be oriented perpendicularly to the plane,
%and $\theta$ is the angle between ${\bf r}$ and the direction of the dipole ($z$).

In order to describe Bose-Bose mixtures at finite temperature,  we divide the Bose-field operator into two parts: 
the condensate contribution $\Phi$, which corresponds to the macroscopic occupation of a single quantum state 
and noncondensed part $\hat {\bar\psi}$, which corresponds to thermally-excited atoms:  
 \begin{equation}\label{FielOp}
\hat \psi_j ({\bf r},t)=\Phi_j({\bf r},t)+\hat {\bar\psi}_j({\bf r},t).
\end{equation}
Within this, the Hamiltonian (\ref{ham})  takes the form of a sum
\begin{equation}\label{ham1}
\hat H=\hat H^{(0)}+\hat H^{(1)}+\hat H^{(2)}+\hat H^{(3)}+\hat H^{(4)},
\end{equation}
 where

%\begin{widetext}
\begin{subequations}\label{sumham}
\begin{align}
\hat H^{(0)} &= \sum_j \bigg [\int d {\bf r } \, \Phi_j^* ({\bf r}) h_j^{sp} \Phi_j ({\bf r}) \label{sumham1} \\ 
&+\frac{1}{2}  \int d {\bf r } \int d {\bf r' }  V_j ({\bf r-r'}) n_{cj} ({\bf r}) n_{cj} ({\bf r'}) \bigg] \nonumber \\
&+  \int d {\bf r } \int d {\bf r' }  V_{12} ({\bf r-r'}) n_{c2} ({\bf r}) n_{c1} ({\bf r'}) \bigg] , \nonumber \\
\hat H^{(1)}&=  0,  \label{sumham2}\\
\hat H^{(2)}&=  \sum_j \bigg \{ \int d {\bf r } \, \hat {\bar\psi}_j^\dagger ({\bf r}) h_j^{sp} \hat {\bar\psi}_j ({\bf r}) \int d {\bf r } \int d {\bf r' }  V_j ({\bf r-r'})   \label{sumham3} \\
& \times \bigg [ n_{cj} ({\bf r}) \hat {\bar\psi}_j^\dagger ({\bf r'}) \hat {\bar\psi}_j ({\bf r'}) + 
\Phi_j^*({\bf r}) \Phi_j({\bf r'}) \hat {\bar\psi}_j^\dagger ({\bf r'}) \hat {\bar\psi}_j ({\bf r})  \nonumber\\
&+ \frac{1}{2} \Phi_j^*({\bf r'}) \Phi^*_j({\bf r}) \hat {\bar\psi}_j ({\bf r'}) \hat {\bar\psi}_j ({\bf r}) 
+ \frac{1}{2} \Phi_j({\bf r'}) \Phi_j({\bf r})  \hat {\bar\psi}_j ^\dagger({\bf r'}) \hat {\bar\psi}_j ^\dagger({\bf r})  \bigg] \bigg\}  \nonumber\\
 &+  \int d {\bf r } \int d {\bf r' } V_{12} ({\bf r-r'})  \bigg [ \hat {\bar\psi}_2^\dagger({\bf r}) \hat {\bar\psi}_2({\bf r}) n_{c1} ({\bf r'})  \nonumber\\
&+\hat {\bar\psi}_1^\dagger({\bf r'}) \hat {\bar\psi}_1({\bf r'}) n_{c2} ({\bf r})  \bigg] , \nonumber\\
\hat H^{(3)}&= \sum_j \bigg\{ \int d {\bf r } \int d {\bf r'} V_j ({\bf r-r'}) \bigg[ \Phi_j ({\bf r}) \hat {\bar\psi}_j^\dagger ({\bf r}) \hat {\bar\psi}_j^\dagger  ({\bf r'}) \hat {\bar\psi}_j  ({\bf r'}) 
\label{sumham4} \\
&+\Phi_j^*({\bf r}) \hat {\bar\psi}_j^\dagger ({\bf r'}) \hat {\bar\psi}_j ({\bf r'}) \hat {\bar\psi}_j ({\bf r}) \bigg]  \bigg \}   \nonumber\\ 
&+\int d {\bf r } \int d {\bf r'} V_{12} ({\bf r-r'}) \bigg[ \Phi_1 ({\bf r'}) \hat {\bar\psi}_2^\dagger ({\bf r}) \hat {\bar\psi}_2 ({\bf r}) \hat {\bar\psi}_1^\dagger  ({\bf r'}) \nonumber\\
&+\Phi_1^*({\bf r'}) \hat {\bar\psi}_2^\dagger ({\bf r}) \hat {\bar\psi}_2 ({\bf r}) \hat {\bar\psi}_1 ({\bf r'})   
+\Phi_2({\bf r}) \hat {\bar\psi}_2^\dagger ({\bf r}) \hat {\bar\psi}_1^\dagger ({\bf r'}) \hat {\bar\psi}_1 ({\bf r'})  \nonumber \\
&+\Phi_2^*({\bf r}) \hat {\bar\psi}_2 ({\bf r}) \hat {\bar\psi}_1^\dagger ({\bf r'}) \hat {\bar\psi}_1 ({\bf r'})\bigg], \nonumber \\
\hat H^{(4)}&= \frac{1}{2} \sum_j \bigg [\int d {\bf r } \int d {\bf r ' }  \hat {\bar\psi}_j^\dagger({\bf r}) \hat {\bar\psi}_j^\dagger ({\bf r'})  \label{sumham5}
V_j ({\bf r-r'}) \hat {\bar\psi}_j  ({\bf r'})\hat {\bar\psi}_j ({\bf r})  \bigg]  \\
&+ \int d {\bf r } \int d {\bf r ' }  \hat {\bar\psi}_1^\dagger({\bf r}) \hat {\bar\psi}_2^\dagger ({\bf r'}) V_{12} ({\bf r-r'}) \hat {\bar\psi}_2  ({\bf r'})\hat {\bar\psi}_1 ({\bf r}).\nonumber
\end{align}
\end{subequations}
%\end{widetext}
In Eqs.(\ref{sumham}), we have used the condition $\langle \hat {\bar\psi}_j \rangle =0$, which ensures the quantum number conservation condition.
At zero temperature $T=0$, almost all of the particles are in the condensate state, hence the noncondensed operator can be neglected ($\hat {\bar\psi}_j=0$), and 
only the zeroth order  $H^{(0)}$ term can be taken into account in Eq.(\ref{ham1}). 
Therefore, the ground state of the system can be described by two coupled Gross-Pitaevskii (GP) equations for the condensate
wavefunctions $\Phi_j({\bf r},t)$.
At finite temperature, the GP equations for Bose-Bose mixtures reads
%the evolution equations for Bose-Bose mixtures can be determined by employing the variational principle 
%of Balian-V\'en\'eroni \cite{BV} which  optimizes a generating functional
%depending on the observables of interest. This yields the equation for the condensates \cite{Boudj, Boudj55, Boudj66, Boudj00}
\begin{align} \label{T:DH}   
i\hbar \dot{\Phi}_j ({\bf r},t) &=\frac{d{\cal E}}{d \Phi_j^*} \nonumber \\
&=  h_j^{sp} \Phi_j ({\bf r},t)+\int d{\bf r'} V_j({\bf r}-{\bf r'}) \bigg [ n_j ({\bf r'},t) \Phi_j({\bf r},t)  \nonumber \\
&+\tilde n_j ({\bf r},{\bf r'},t)\Phi_j ({\bf r'},t) +\tilde m_j ({\bf r},{\bf r'},t)\phi_j^*({\bf r'},t) \bigg ] \nonumber \\
&+\int d{\bf r'} V_{12} ({\bf r}-{\bf r'}) n_{3-j} ({\bf r'}) \Phi_j ({\bf r},t),  
\end{align}
where  ${\cal E}=\langle \hat H\rangle$ is the energy of the system, and
$n_{cj}({\bf r})=|\Phi_j({\bf r})|^2$, $\tilde n_j ({\bf r})= \langle \hat {\bar\psi}_j^\dagger ({\bf r}) \hat {\bar\psi}_j ({\bf r}) \rangle $ and 
$\tilde m_j ({\bf r})= \langle \hat {\bar\psi}_j ({\bf r}) \hat {\bar\psi}_j ({\bf r}) \rangle $ are, respectively  the condensed, noncondensed and anomalous densities.
The total density in each components is given by $n_j({\bf r})=n_{cj}({\bf r})+\tilde n_j ({\bf r})$.
The quantities $\tilde n_j ({\bf r, r'})$ and $\tilde m_j ({\bf r, r'})$ stand for the normal and the anomalous one-body density matrices
which account for the dipole exchange interaction between the condensate and noncondensate. 
The total number of particles is defined as $N_j=N_{cj}+\tilde N_j=\int n_j d {\bf r}$, where  
$N_{cj}=\int n_{cj} d {\bf r}$ and $\tilde N_j=\int \tilde n_j d {\bf r}$ are respectively, the condensed and the noncondensed number of particles in each component.
For $r_{*1}=r_{*2}=r_{*12}=0$, the coupled GP equations (\ref{T:DH}) reduce to those of a finite-temperature nondipolar mixture \cite{Boudj00}.
If $\tilde n_j=\tilde m_j=0$, one can reproduce the usual GP equations for binary condensates at zero temperature.
The dynamics of the noncondensed and the anomalous densities can be derived easily using the coupled TDHFB equations
 \cite{Boudj, Boudj55, Boudj66, Boudj00}.

In what follows we consider only mixtures with equal mass.
In the uniform case, for which the trapping  potentia vanishes ($U_j=0$), translational invariance requires the solutions to be plane waves.
The noncondensed operators can be written as  $\hat {\bar\psi}^{\dagger}_j ({\bf r})= (1/V) \sum_{\bf k} \hat a^{\dagger}_{j \bf k} e^{-i \bf k. \bf r}$
and $\hat {\bar\psi}_j ({\bf r})= (1/V) \sum_{\bf k} \hat a_{j \bf k} e^{i \bf k. \bf r}$, where  
$\hat a_{\bf k}^\dagger$ and $\hat a_{\bf k}$ are, respectively the creation and annihilation operators of particles and  $V$ is a quantization volume.
The Fourier transforms of interaction potentials (\ref {IntraPot})  and (\ref {InterPot}) are given by 
\begin{align}
\tilde V_j(\mathbf k)&=g_j [1+\epsilon_j^{dd} (3\cos^2\theta_k-1)], \\
\tilde V_{12}(\mathbf k)&=g_{12} [1+\epsilon_{12}^{dd} (3\cos^2\theta_k-1)],
\end{align}
 where $\epsilon_j^{dd}=r_{*j}/3a_j$ and $\epsilon_{12}^{dd}=r_{*12}/3a_{12}$.

After having simplifying the higher-order terms (\ref{sumham4}) and (\ref{sumham5}) applying the HFB approximation, 
the resulting Hamiltonian can be diagonalized by employing the following canonical Bogoliubov transformations \cite{Tom}: 
\begin{subequations} \label{T:B}
\begin{align} 
 \hat a_{1k}&= (u_{1k} \hat b_{1k}+ v_{1k} \hat b_{1,-k}^\dagger) \cos \gamma
- (u_{2k} \hat b_{2k}+ v_{2k} \hat b_{2,-k}^\dagger) \sin \gamma,  \\ 
 \hat a_{2k}&= (u_{1k} \hat b_{1k}+ v_{1k} \hat b_{1,-k}^\dagger) \sin \gamma  
+ (u_{2k} \hat b_{2k}+ v_{2k} \hat b_{2,-k}^\dagger) \cos \gamma, 
\end{align}
\end{subequations}
where $\hat b_k$ and $ \hat b_k^\dagger$ are the quasi-particle  operators satisfying the usual Bose commutation relations, 
the Bogoliubov  functions $ u_{jk}$ and $v_{jk}$ are given by 
\begin{equation} 
 u_{jk},v_{jk}= \frac{1}{2}\left(\sqrt{\varepsilon_{jk}/E_k}\pm\sqrt{E_k/\varepsilon_{jk}} \right),
\end{equation}
where $E_k=\hbar^2k^2/2m$ is the kinetic energy of a particle and $\varepsilon_{jk}$ is the Bogoliubov excitations energy.\\
Keeping in mind that the Bogoliubov approximation is valid only for asymptotically weak interactions and at very low temperatures where
$\tilde n_j \ll n_{cj}$ and $\tilde m_j \ll n_{cj}$. 
Indeed, this is equivalent to the case where  the long-range exchange term  $\tilde n ({\bf r},{\bf r'})=\tilde m ({\bf r},{\bf r'})=0$ \cite{Bon}
which does not influence the stability of the system \cite {Bon, BoudjDp, Boudj2017}.
This condition is necessary to guarantee the gaplessness of  the spectrum i.e. $\lim\limits_{k \rightarrow 0} \varepsilon_{jk}=0$, 
and the Hugenholtz-Pines \cite{HP} theorem.  
Within this we obtain for the Bogoliubov spectrum
\begin{equation} \label {Bog}
\varepsilon_{1k}= \sqrt{E_k^2+2E_k \nu_1(\theta)},  \,\,\,\, \varepsilon_{2k}= \sqrt{E_k^2+2E_k \nu_2(\theta)}, 
\end{equation}
where
\begin{equation} \label {Chmp}
 \nu_{1,2}(\theta)=  \frac{\tilde V_1({\bf k}) n_{c1}} {2} f_{1,2}(\theta),
\end{equation}
$$f_{1,2}= 1 + \alpha \pm \sqrt{ (1-\alpha)^2 +4 \Delta ^{-1}\alpha },$$
and 
\begin{align} 
\cos \gamma, \sin \gamma= \frac{1}{\sqrt{2}}  \sqrt{1\pm \frac { 1-\alpha}  { \sqrt{ (1-\alpha)^2 +4\Delta^{-1} \alpha}}}, 
\end{align}
where 
$$\alpha(\theta)=  \beta \frac{1+\epsilon_2^{dd} (3\cos^2\theta-1)}{1+\epsilon_1^{dd} (3\cos^2\theta-1)},$$
with  $\beta= n_{c2} g_2 /n_{c1} g_1$. \\  
The miscibility parameter is defined as
\begin{align} \label {mc}
\Delta(\theta)&=\frac{\tilde V_1({\bf k}) \tilde V_2({\bf k})} { \tilde V_{12}^2({\bf k})} \\
& =\Delta \frac{ [1+\epsilon_1^{dd} (3\cos^2\theta-1)] [1+\epsilon_2^{dd} (3\cos^2\theta-1)] } {[1+\epsilon_{12}^{dd} (3\cos^2\theta-1)]^2}, \nonumber
\end{align} 
where $\Delta=g_1 g_2/g_{12}^2$ is the miscibility parameter of a nondipolar mixture. 
For $\Delta (\theta) >1$, the mixture is in a stable miscible regime, while $\Delta (\theta)<1$ leads to an unstable immiscible phase 
for any value of $\theta$ \cite {Wilson}. 
The IMT occurs when the interspecies and intraspecies are balanced i.e $\Delta(\theta)=1$.
For $\theta=\pi/2$, i.e. when momenta are perpendicular to the dipole direction, 
$\Delta (\pi/2) =\Delta  (1-\epsilon_1^{dd}) (1-\epsilon_2^{dd}) /(1-\epsilon_{12}^{dd})^2$. In such a situation,
a stable mixture requires the condition $\epsilon_j^{dd} = 1+ \left[ (1-\epsilon_{12}^{dd})^2/ \Delta (1-\epsilon_{3-j}^{dd}) \right] \geq 1$.
For $\epsilon_1^{dd}=\epsilon_2^{dd}=\epsilon_{12}^{dd}=0$, the miscibility parameter becomes $\Delta (\theta) \equiv \Delta$.

In the long-wavelength limit ($k \rightarrow 0$), the Bogoliubov excitations (\ref{Bog}) are sound waves $\varepsilon_{jk}= \hbar c_j (\theta) k$,  
where $c_j (\theta)= \sqrt{\tilde V_j(|{\bf k} |=0)  n_{cj} /m_j}$ is the sound velocity of a single condensate. 
As a result, the total dispersion is phonon-like 
\begin{equation} \label{sound}
\varepsilon_{1,2 k}= \hbar c_{1,2} (\theta) k,
\end{equation} 
where the sound velocities $c_{1,2}$  are
 \begin{equation} \label{sound1}
c_{1,2} ^2 (\theta)=\frac{1}{2} \left[ c_1^2+c_2^2 \pm \sqrt{ \left( c_1^2-c_2^2\right) ^2 + 4 \Delta^{-1} c_1^2 c_2^2} \right] (\theta).
\end{equation}
For $\Delta  (\theta) >1$, $c_2$ tends to zero, indicating  that the system becomes unstable and thus, the two condensates spatially separate
in agreement with our above predictions.
Remarkably, the sound velocity is angular dependence; in other words, it acquires a dependence on
the propagation direction, $\theta$ owing to the anisotropy of the DDI. 
In the case of a single dipolar BEC, the anisotropy of the sound velocity has already been observed experimentally in Ref.\cite{bism}.

The diagonalized Hamiltonian reads 
\begin{equation} \label {diagH}
\hat H = E+ \sum \limits_{j=1}^2 \sum \limits _{\bf k} \varepsilon_{j k} \hat b_{j \bf k}^\dagger \hat b_{j\bf k},
\end{equation}
where  $E=E_0+ \delta E$ is the ground state energy of the system with
$$E_0= \frac{1}{2}  \sum\limits _{j=1}^2 \tilde V_j( |{\bf k}|=0) n_{cj}^2 + \tilde V_{12} (|{\bf k}|=0) n_{c1}  n_{c2},$$
is anisotropic and should be evaluated at $k \rightarrow 0 $ since it accounts for the condensate \cite{lime, Boudj2015}. \\
And
\begin{align}  \label {LHYEgy}
\delta E&= \frac{1}{2} \sum\limits_{j=1}^2 \sum\limits_k \left [\varepsilon_{jk} - E_k - n_{cj} \tilde V_j({\bf k}) +\frac{ n_{cj} ^2\tilde V^2_j({\bf k})}{2E_k} \right]  \nonumber \\
&+\frac{1}{2}  \sum\limits_k \frac{ n_{c1} n_{c2} \tilde V^2_{12}({\bf k})}{E_k}, 
\end{align}
stands for the ground-state energy corrections due to quantum fluctuations \cite{Petrov,Capp, Larsen, Boudj00}. 
The last two terms in Eq.(\ref{LHYEgy}) have been added in order to circumvent the ultraviolet divergence arising in that integrals.

\section{Fluctuations and thermodynamics} \label{Flth}

Explicit expressions for the noncondensed  density $\tilde n_j =  \sum\limits_{k\neq 0}  \langle  \hat a^\dagger_{jk} \hat a_{jk}\rangle$
and the anomalous density $\tilde m_j =  \sum\limits_{k\neq 0}  \langle  \hat a_{jk} \hat a_{jk}\rangle$ (density of pair-correlated atoms)
can be given by utilizing the transformation (\ref{T:B}).  This yields
\begin{align} 
 \tilde n_j &= \frac{1}{V}  \sum\limits_{k\neq 0} \bigg\{ \bigg[ v_{jk}^2+ \left (u_{jk}^2+v_{jk}^2 \right) N_{jk}\bigg] \cos^2 \gamma  \label {HFB1}  \\ 
&+\bigg[v_{(3-j)k}^2+ \left (u_{(3-j)k}^2+v_{(3-j)k}^2 \right) N_{(3-j)k}\bigg]  \sin^2 \gamma \bigg \},  \nonumber \\
 \tilde m_j&= -\frac{1}{V}  \sum\limits_{k\neq 0} \bigg\{ \left[ u_{jk} v_{jk} (1+2N_{jk}) \right] \cos^2 \gamma  \label {HFB2}  \\ 
&+ \left[ u_{(3-j)k} v_{(3-j)k} (1+2N_{(3-j)k} )\right] \sin^2 \gamma \bigg\}, \nonumber 
\end{align}
where $N_{jk}=\langle \hat b_{jk}^{\dagger} \hat b_{jk}\rangle=[\exp(\varepsilon_{jk}/T)-1]^{-1}$ are occupation numbers for the excitations.  

In the thermodynamic limit,  the discrete sum over $k$ can be replaced by an integral over a continuous variable $k$ as follows:
$(1/V)\sum_k=\int d {\bf k}/ (2\pi)^3$.  Therefore, we obtain for the condensed depletion 

\begin{align}\label {NCD}
\tilde n_j&=  \frac{1}{2\sqrt{2}}\tilde n_1^0 \left[ {\cal I}_j^3 (\epsilon_{dd}) + {\cal I}_{3-j}^3 (\epsilon_{dd}) \right] \\
&+2\sqrt{2}\,\tilde n_1^{th} \left[ {\cal I}_j^{-1} (\epsilon_{dd}) + {\cal I}_{3-j}^{-1} (\epsilon_{dd}) \right], \nonumber
\end{align}
where $\tilde n_1^0= (8/3) n_{c1}  \sqrt{ n_{c1} a_1^3/\pi} $ is the zero temperature single condensate depletion (type-1)
and $\tilde n_1^{th}=(2/3) n_{c1} \sqrt{ n_{c1} a^3/\pi} (\pi T/n_{c1} g)^2$ is the thermal contribution to the noncondensed density of a single condensate. 
The functions ${\cal I}_j^{\ell} (\epsilon_{dd})$, which are defined as 
\begin{subequations}\label {Nfunc}
\begin{align} 
{\cal I}_j^{\ell} (\epsilon_{dd})&= \int_0^{\pi}  \sin \theta \left[1+\epsilon_1^{dd} (3\cos^2\theta-1) \right]^{\ell/2}  \\
& \times f_j^{\ell/2} \cos^2\gamma \, d \theta, \nonumber \\
{\cal I}_{3-j}^{\ell} (\epsilon_{dd})&= \int_0^{\pi}  \sin \theta \left[1+\epsilon_1^{dd} (3\cos^2\theta-1) \right]^{\ell/2}  \\
& \times  f_{3-j}^{\ell/2} \sin^2\gamma \, d \theta, \nonumber
\end{align}
\end{subequations}
account for the DDI  contribution to the condensate depletion.
At zero temperature and for $\epsilon_1^{dd}=\epsilon_2^{dd}= \epsilon_{12}^{dd}=0$,  the depletion (\ref{NCD}) reduces  
to that obtained in Refs. \cite{Tom, SSS} using the Bogoliubov theory.

As is clearly seen in Eq.(\ref{HFB2}), the expression of $\tilde m$ is ultraviolet divergent.
This inconsistency is a symptom of the contact interaction. 
Indeed, there are many ways of dealing with such a problem, for instance, the dimensional regularization 
which gives for the integral $\int_0^{\infty} dx (x/ \sqrt{1+x^2})=-1$ \cite{Anders, Yuk, Boudj}.
We proceed along the lines of Ref \cite{Boudj00}, and obtain the following for the anomalous density 
\begin{align} \label {AND} 
\tilde m_j&= \frac{1}{2\sqrt{2}} \tilde m_1^0 \left[ {\cal I}_j^3 (\epsilon_{dd}) + {\cal I}_{3-j}^3 (\epsilon_{dd}) \right]\\
&+ 2\sqrt{2} \,\tilde m_1^{th} \left[ {\cal I}_j^{-1} (\epsilon_{dd}) + {\cal I}_{3-j}^{-1} (\epsilon_{dd}) \right], \nonumber
\end{align}
where $\tilde m_1^0= 8 n_{c1}  \sqrt{ n_{c1} a_1^3/\pi} $ is the anomalous density of the single component at zero temperature
and $\tilde m_1^{th}=-\tilde n_1^{th}$ is the thermal contribution to the  anomalous density of a single condensate. 
In fact, the resulting pair anomalous correlation is important since it provides insights
into the phenomenon of dissipation and superfluidity (see below). 
Moreover, such a quantity might give hints about the superradiance in dipolar ultracold atoms. 

The leading term in Eqs.(\ref{NCD}) and (\ref{AND}) stands for the quantum fluctuations. 
The subleading term which represents the thermal fluctuations, is evaluated at temperatures $T\ll g n_c $, 
where the main contribution to Eqs.(\ref{HFB1}) and (\ref{HFB2}) comes from the phonon branch.
At temperatures $T \gg g n_c$, the main contribution to (\ref{HFB1}) comes from the single-particle excitations. 
Therefore, the thermal contribution of $\tilde n$ becomes identical to the density of noncondensed atoms in an ideal Bose gas 
while the pair anomalous correlation cannot exist any more in such a regime. 
In the absence of the DDI, Eqs.(\ref{NCD}) and (\ref{AND})  excellently agree with our equations obtained recently for a nondipolar mixture \cite{Boudj00}. 
For $\epsilon_{12}^{dd}=0$, expressions of $\tilde m_j$ and $\tilde n_j$ reduce to those found for a single dipolar BEC \cite {Boudj2015,Boudj2jpa}.
The comparaison between Eqs.(\ref{NCD}) and (\ref{AND}) reveals that the anomalous correlation is always greater than the condensate depletion as in the case of a single BEC.
Both quantities are monotonically increasing with $\epsilon_{dd}$. 
We see also that the effects due to quantum fluctuations are small compared to those due to thermal fluctuations
since ${\cal I}_j^{-1} (\epsilon_{dd}) > {\cal I}_j^{3} (\epsilon_{dd})$. 
For instance, for $\beta = 0.2$, $\Delta = 1.5$, and  $\epsilon_1^{dd} =\epsilon_2^{dd} =\epsilon_{12}^{dd} \simeq 1$, 
the quantum depletion is larger by $\sim 5.55$ than that of a single-component BEC with contact interactions,  
whereas at finite temperature,  the thermal depletion in each component is about 18 times higher than that of one Bose gas.
These values are decreasing with increasing both $\beta$ and $\Delta$ and become imaginary for $\Delta>1$ signaling that the system is unstable. 
The same behavior holds for the pair anomalous correlations.
%The total noncondensed and anomalous densities are, respectively  $\tilde n = \sum_j \tilde n_j $ 
%and  $\tilde m = \sum_j \tilde m_j $.  

The Bogoliubov approach requires that quantum and thermal fluctuations should be small.
Therefore, the small parameter of the theory can be given as $\sqrt{ n_{c1} a_1^3} \,[ {\cal I}_1^3 (\epsilon_{dd})+ {\cal I}_2^3 (\epsilon_{dd})]  \ll 1$
and $(T/ n_{c1} g_1) \sqrt{ n_{c1} a_1^3} \, [ {\cal I}_1^{-1} (\epsilon_{dd})+ {\cal I}_2^{-1} (\epsilon_{dd})]  \ll 1$.
In the absence of the DDI and the interspecies interaction, the validity criterion of the theory reduces to $ \sqrt{ n_c a^3} \ll 1$.

%The corrections to the ground state energy due to quantum fluctuations can be calculated via (\ref{LHYEgy}) as :
%\begin{equation} \label{GSE}
%\delta E_j =\frac{1}{4\sqrt{2}}  E_1^0  \left[ {\cal I}_j^5 (\epsilon_{dd}) + {\cal I}_{3-j}^5 (\epsilon_{dd}) \right],
%\end{equation}
%where $E_1^0/V= (64/15) g_1 n_{c1}^2 \sqrt{ n_{c1} a^3/\pi}$ is the ground state energy of a single BEC.
%If $\epsilon_{12}^{dd}=0$, $\delta E$ reproduces the ground state energy of a single dipolar Bose gas \cite{lime, Boudj2015}. 
%For $\epsilon_1^{dd}=\epsilon_2^{dd}= \epsilon_{12}^{dd}=0$, one recovers the Larsan's formula \cite{Larsen}.

The shift to the equation of state (EoS) due to quantum and thermal fluctuations can be obtained through 
$\delta \mu_j=\sum\limits_{\bf k} \tilde V({\bf k}) [v_{jk}(v_{jk}-u_{jk})]=\sum\limits_{\bf k} \tilde V({\bf k}) (\tilde{n}_j+\tilde{m}_j)$ 
\cite{BoudjG, Boudj2015, Boudj00, Boudj3}.
\begin{align} \label{EoS}
\delta \mu_j& = \frac{1}{4\sqrt{2}}  \mu_1^0  \left[ {\cal I}_j^5 (\epsilon_{dd}) f_j^{-1/2}+ {\cal I}_{3-j}^5 (\epsilon_{dd}) f_j^{-1/2}\right] \\
& + \frac{ \sqrt{2}\, m T^2} {12 \hbar^3}  \left[ \frac{ {\cal I}_j^1 (\epsilon_{dd}) f_j^{-1}}{c_1}+ \frac{ {\cal I}_{3-j}^1 (\epsilon_{dd}) f_j^{-1}}{c_2} \right], \nonumber
\end{align}  
where $\mu_1^0= (32/3) g_1 n_{c1} \sqrt{n_{c1} a_1^3/\pi}$ is the EoS of a single Bose gas.
For $\epsilon_1^{dd}=\epsilon_2^{dd}= \epsilon_{12}^{dd}=0$, and $g_{12}$=0, Eq.(\ref{EoS}) recovers the celebrated LHY corrections
of the chemical potential \cite{LHY} for  a single Bose gas.

At $T=0$, the inverse compressibility is equal to $ \kappa_j^{-1}=n_j^2\partial\mu_j/\partial n_j$. 
Then the quantum corrections to the inverse compressibility matrix can be computed via Eq.(\ref{EoS}), and we obtain 
\begin{align}  \label{comp}
\frac{\partial \delta \mu_1} {g_1\partial n_1} &= \frac{4}{\sqrt{2}} \sqrt{ \frac{n_{c1} a_1^3}{\pi}} 
\bigg[  {\cal G}_1(\theta) f_1^{-3/2} {\cal I}_1^5 (\epsilon_{dd}) + f_1^{-1/2} {\cal I}_1^5 (\epsilon_{dd})  \nonumber\\
&- {\cal G}_2(\theta) f_2^{-3/2} {\cal I}_2^5 (\epsilon_{dd}) + f_2^{-1/2} {\cal I}_2^5 (\epsilon_{dd}) \bigg],
\end{align}
and 
\begin{align}  \label{comp1}
\frac{\partial \delta \mu_2} {g_2\partial n_2} &= \frac{4}{\sqrt{2}} \sqrt{ \frac{n_{c1} a_1^3}{\pi}} 
\frac{1}{\beta}   \bigg[- {\cal G}_1(\theta) f_1^{-3/2} {\cal I}_1^5 (\epsilon_{dd}) \nonumber\\
&+ {\cal G}_2(\theta) f_2^{-3/2} {\cal I}_2^5 (\epsilon_{dd}) \bigg],
\end{align}
where  $${\cal G}_{1,2}= \mp \alpha+ \frac{ \alpha(1-\alpha)-2\alpha \Delta ^{-1}} {\sqrt{(1-\alpha)^2+4\alpha \Delta ^{-1}}}.$$

At finite temperature, the grand-canonical ground state energy can be calculated using  the thermodynamic relation
$E'=E+E^{th}=-T^2 \left( \frac{\partial} {\partial T } \frac{F}{T} \right) |_{V,\mu}$ where the free energy is given by 
$F=E'+T\sum_{\bf k}\ln[1-\exp(-\varepsilon_{kj}/T)]$.
At low temperature $T\ll g_1 n_{c1}$, corrections to the ground state energy due to thermal fluctuations are 
\begin{align} \label{GSEth}
E^{th}= \frac{ \sqrt{2}\pi^2 T^4} {15 \hbar^3}  \left[  \frac{ {\cal J}_1^{-3} (\epsilon_{dd})}{c_1^3} + \frac{ {\cal J}_2^{-3} (\epsilon_{dd})} {c_2^3} \right],
\end{align}
where the functions $ {\cal J}_j^{\ell}  (\epsilon_{dd}) $ are defined as 
\begin{equation} \label {Nfunc1}
{\cal J}_j^{\ell} (\epsilon_{dd})= \int_0^{\pi}  \sin \theta \left[1+\epsilon_1^{dd} (3\cos^2\theta-1) \right]^{\ell/2}  f_j^{\ell/2} \, d \theta, 
\end{equation} 

The system pressure is defined as $P=-(\partial F/\partial V)|_T$. 
Again at $T\ll g_1 n_{c1}$, the thermal pressure is 
\begin{equation}\label {therpress}
P^{th}= \frac{ \sqrt{2}\pi^2 T^4} {45 \hbar^3}  \left[  \frac{ {\cal J}_1^{-3} (\epsilon_{dd})}{c_1^3} + \frac{ {\cal J}_2^{-3} (\epsilon_{dd})} {c_2^3} \right].
\end{equation}
The explicit value of $P^{th}$ enables us to estimate the inverse isothermal compressibility of the gas  
$ (\partial P_j^{th}/\partial n_j)$.
%\begin{equation}\label {Isocomp}
%\left(\frac{\partial P} {\partial n}\right)^{th}= -\frac{\sqrt{2} \pi^2T^4}{30 m\hbar^3}  
%\left[  \frac{ {\cal J}_1^{-3} (\epsilon_{dd})}{c_1^5} + \frac{ {\cal J}_2^{-3} (\epsilon_{dd})} {c_2^5} \right].
%\end{equation}

Remarkably, corrections due to quantum and thermal fluctuations to all the above thermodynamic quantities are 
isotropic while their leading terms are anisotropic i.e. their values depend on the propagation direction.
For vanishing DDIs, Eqs.(\ref{EoS})-(\ref{therpress}) reduce to those found for a nondipolar mixture \cite{Boudj00}. 

It is interesting to discuss the case of a balanced mixture where $n_{c1} = n_{c2}$ and $\tilde V_1({\bf k}) =\tilde V_2({\bf k})= \tilde V_{12}  ({\bf k})= \tilde V  ({\bf k})$, one has 
$f_1=4$ and $f_2=0$ and  hence,  the spectrum of the upper branch is identical to the spectrum of the one-component dipolar BEC, 
$\varepsilon_{1k}= \sqrt{E_k^2+8 E_k n_{c1} V({\bf k})}$.
In such a case the functions (\ref{Nfunc}) reduce to  
${\cal I}_j^{\ell} (\epsilon_{dd})={\cal Q}_{\ell} (x)=(1-x)^{\ell/2} {}_2\!F_1\left(-\frac{\ell}{2},\frac{1}{2};\frac{3}{2};\frac{3x}{x-1}\right)$, 
where ${}_2\!F_1$ is the hypergeometric function. The functions ${\cal Q}_{\ell} (x)$ are maximum at $x \approx 1$. 
For $x \geq1$,  ${\cal Q}_{\ell} (x)$ are imaginary which means that the dipolar interaction dominates the repulsive two-body 
interactions leading to unstable soft modes, 
whereas, the spectrum associated with the lower branch becomes identical to that of free particles, $\varepsilon_{2k}=E_k$.
Therefore, the noncondensed and the anomalous densities take the following forms:
\begin{equation}\label {NCD1}
\tilde n= 2\sqrt{2} \, \tilde n_1^0 Q_3 (\epsilon_{dd})+\sqrt{2}\,\tilde n_1^{th} Q_{-1} (\epsilon_{dd})+ \Lambda^3 \zeta(3/2),
\end{equation}
and 
\begin{equation}\label {AND1} 
\tilde m= 2\sqrt{2} \, \tilde m_1^0 Q_3 (\epsilon_{dd}) +\sqrt{2}\,\tilde m_1^{th} Q_{-1} (\epsilon_{dd}),
\end{equation}
where $\Lambda$ is the thermal de-Broglie wavelength and $\zeta (3/2)$ is the Riemann zeta function.
The anomalous density cannot exist in the component associated with the lower branch since the system becomes extremely dilute.
We see that the thermal term in a balanced mixture is larger by $\sqrt{2}$ than that of a single BEC \cite{lime, Boudj2015}.

The ground-state energy simplifies to  
\begin{equation} \label{eeg}
\delta E=(8/\sqrt{2}) E_1^0  Q_5 (\epsilon_{dd}),
\end{equation}
where $E_1^0/V= (64/15) g_1 n_{c1}^2 \sqrt{ n_{c1} a_1^3/\pi}$ is the zero-temperature single condensate ground state energy.
The energy (\ref{eeg}) differs by the factor $8/\sqrt{2}$ from the one Bose gas \cite{lime, Boudj2015}.
For $\epsilon_{dd} >1$,  the DDIs would destabilize the balanced mixture due to the possibility of the collapse.  
%However, as in the case of a single component \cite {BoudjDp,Pfau, Wach, Bess, Chom}, the repulsive LHY corrections may arrest the dipolar attraction forces 
%leading to the formation of a stable long-range quantum mixture droplet. 

Now let us look at the situation where one component is dipolar and the other is nondipolar, say component 2. 
This leads to  $r_{*2}=r_{*12}=r_{*21}=0$ and thus,
$\alpha(\theta)=\beta [1+\epsilon_1^{dd} (3\cos^2\theta-1)]^{-1}$.   
The miscibility parameter (\ref{mc}) takes the form $\Delta(\theta)=\Delta [1+\epsilon_1^{dd} (3\cos^2\theta-1)]$.
In the vicinity of phase separation transition, one has $\tilde V_1({\bf k})  \tilde V_2({\bf k})  \rightarrow \tilde V_{12}^2({\bf k})$, and 
\begin{align} 
\nu_1 (\theta)& \simeq  \tilde V_1({\bf k}) n_{c1} +  \tilde V_2({\bf k}) n_{c2} , \nonumber \\
\nu_2 (\theta)& \simeq  \frac{ \tilde V_1({\bf k}) n_{c1} \tilde V_2({\bf k}) n_{c2}-\tilde V_{12}^2({\bf k}) n_{c1} n_{c2}} {\tilde V_1({\bf k}) n_{c1} +\tilde V_2({\bf k}) n_{c2} } 
\ll \nu_1 (\theta). \nonumber
\end{align}
Therefore, at zero temperature the noncondensed and anomalous densities become
\begin{equation} \label {NCD2}
\tilde n= \tilde n_1^0 (1+\beta)^{3/2} Q_3 (\varsigma_{dd}), 
\end{equation}
and 
\begin{equation} \label {AND2}
\tilde m= \tilde m_1^0 (1+\beta)^{3/2} Q_3 (\varsigma_{dd}), 
\end{equation}
where $\varsigma_{dd}=\epsilon_{dd}/ (1+ \beta)$. 
A similar formula for the depletion (\ref{NCD2}) has been obtained recently in \cite{Past} using the one-loop approximation.

From Eqs.(\ref{NCD2}) and (\ref{AND2}) follows a useful LHY corrected EoS 
\begin{align}\label {chem2}
\delta \mu =  \mu_1^0 (1+ \beta)^{5/2} Q_5 (\varsigma_{dd}).
\end{align}
The ground-state energy shift can be immediately  calculated via $\delta E = \int \delta \mu \, dn$ as:
\begin{align}\label {eeg2}
\delta E =  E_1^0 (1+ \beta)^{5/2} Q_5 (\varsigma_{dd}).
\end{align}

Near the phase separation  and at low temperature, the lower branch has the free-particle dispersion law $\varepsilon_{k2} = E_k$ \cite {CFetter},  
while the upper branch is phonon-like $\varepsilon_{k1} =\hbar c_1 (1 + \alpha)^{1/2} k$. 
This results indicates that the thermal depletion has two different temperature dependence forms: $\tilde n^{th}=a T^2+ bT^{3/2}$, where 
$a=\tilde n_1^{th} (1 + \beta)^{-1/2} Q_{-1} (\varsigma_{dd})$, and $b= (m /2\pi \hbar^2)^{3/2} \zeta (3/2)$. 
One can conclude that the component related to the lower branch is extremely dilute. 
Notice that the distinction in the temperature dependence cannot hold in the thermal part of the anomalous density 
because $\tilde m^{th}$ itself cannot survive any more in the free particle regime.
All the thermodynamic quantities can be straightforwardly calculated following the procedure outlined in Sec.\ref{Flth}.
For $\epsilon_{dd}> (1+ \beta)$, the mixture becomes unstable even in the miscible regime ($\Delta>1$)
since the function $Q_5 (\varsigma_{dd})$ is imaginary. 
In this case, the repulsive two-body contact interactions are dominated by 
the attractive DDIs, driving the system collapse results in from the presence of unstable soft modes.

\section{Inhomogeneous mixture} \label{InhM}

\begin{figure}
\centerline{
\includegraphics[scale=0.46]{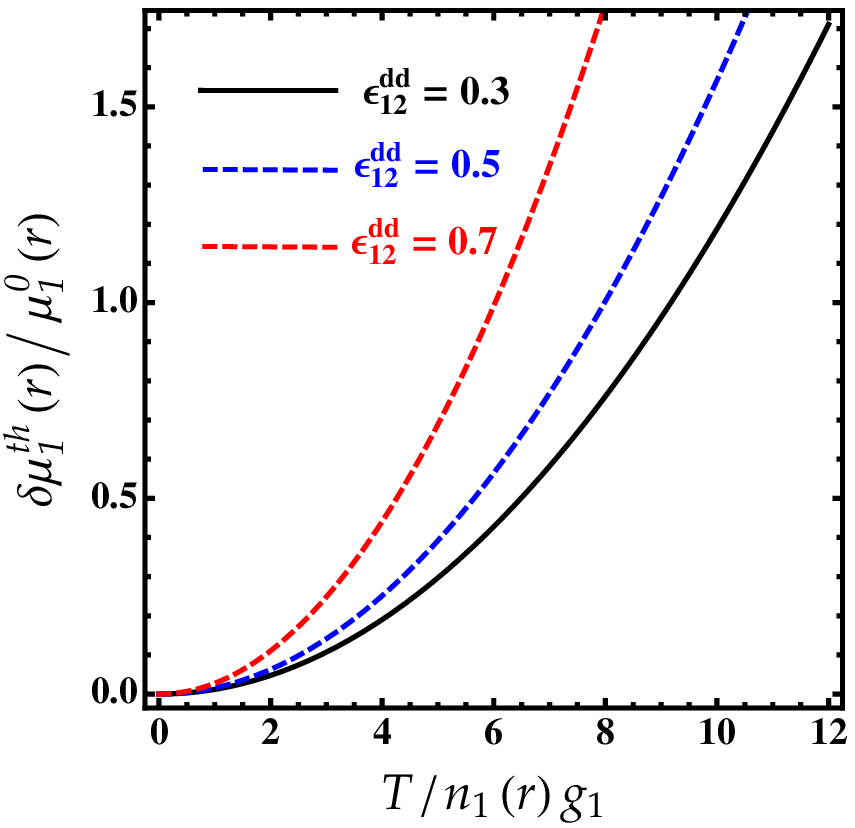}
\includegraphics[scale=0.46]{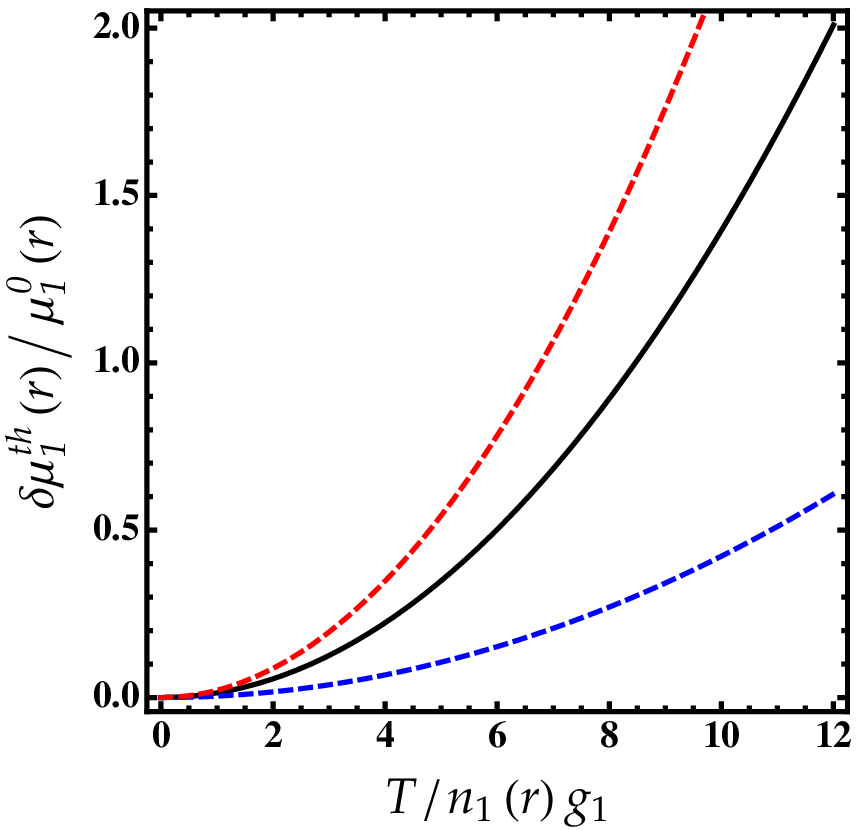} }
 \caption{ Thermal local LHY  corrections as a function of $T/n_1(r) g_1$ for several values of $\epsilon_{12}^{dd}$.
 Parameters are as follows: $\beta = \Delta=1$, and $\epsilon_1^{dd}=\epsilon_2^{dd}=0.7$ (left).  $\beta=1.8$, $\Delta=0.85$, and $\epsilon_1^{dd}=\epsilon_2^{dd}=0.7$ (right). }
\label{TLHY} 
\end{figure}

Now we discuss the case of a harmonically trapped dipolar Bose-Bose mixture, 
$U_j({\bf r})=\frac{1}{2}m (\omega_{jx}^2  x^2+ \omega_{jy}^2 y^2 + \omega_{jz}^2 z^2)$,
where $\omega_{j x,y,z}$ are the trapping frequencies.
To calculate the excitations spectra and the fluctuations, we employ the LDA introduced first for cleaned \cite{lime} 
and disordered \cite{Boudj4, Boudj5} single component dipolar Bose gases.
Such an approximation is valid when the external trapping potentials are sufficiently smooth and hence, consists of setting
$ \varepsilon_k \rightarrow \varepsilon_k ({\bf r})$, and $ u_{jk}\rightarrow  u_{jk}({\bf r})$ and $v_{jk}\rightarrow  v_{jk}({\bf r})$.
Therefore, the noncondensed and the anomalous densities become
\begin{align}\label {NCDih}
\tilde n_j ({\bf r}) &=  \frac{1}{2\sqrt{2}}\tilde n_1^0 ({\bf r}) \left[ {\cal I}_j^3 ({\bf r},\epsilon_{dd}) + {\cal I}_{3-j}^3 ( {\bf r}, \epsilon_{dd}) \right] \\
&+2\sqrt{2}\,\tilde n_1^{th} ({\bf r}) \left[ {\cal I}_j^{-1} ({\bf r},\epsilon_{dd}) + {\cal I}_{3-j}^{-1} ({\bf r}, \epsilon_{dd}) \right], \nonumber
\end{align}
and 
\begin{align}\label {ANDih}
\tilde m_j ({\bf r}) &=  \frac{1}{2\sqrt{2}}\tilde m_1^0 ({\bf r}) \left[ {\cal I}_j^3 ({\bf r},\epsilon_{dd}) + {\cal I}_{3-j}^3 ( {\bf r}, \epsilon_{dd}) \right] \\
&+2\sqrt{2}\,\tilde m_1^{th} ({\bf r}) \left[ {\cal I}_j^{-1} ({\bf r},\epsilon_{dd}) + {\cal I}_{3-j}^{-1} ({\bf r}, \epsilon_{dd}) \right]. \nonumber
\end{align}
The EoS turns out to be given as  $\delta \mu_j ({\bf r})=\delta \mu_j^0 ({\bf r})+\delta \mu_j^{th} ({\bf r})$, where 
\begin{align} \label{EoSih}
\delta \mu_j^0 ({\bf r})& = \frac{1}{4\sqrt{2}}  \mu_1^0 ({\bf r}) \bigg[ {\cal I}_j^5 ({\bf r}, \epsilon_{dd}) f_j^{-1/2} ({\bf r}) \\
&+ {\cal I}_{3-j}^5 ({\bf r},\epsilon_{dd}) f_j^{-1/2}({\bf r})\bigg],   \nonumber 
\end{align}
and 
\begin{align} \label{EoSihT}
\delta \mu_j^{th} ({\bf r})& = \frac{ \sqrt{2}\, m T^2} {12 \hbar^3}  \bigg[ \frac{ {\cal I}_j^1 ({\bf r}, \epsilon_{dd}) f_j^{-1}  ({\bf r})}{c_1 ({\bf r})}  \\
&+\frac{ {\cal I}_{3-j}^1 ({\bf r}, \epsilon_{dd}) f_j^{-1} ({\bf r}) }{c_2  ({\bf r})} \bigg]. \nonumber
\end{align}
The behavior of $\delta \mu_1^{th} ({\bf r})$ is displayed in Fig.\ref{TLHY}. 
It is monotonically increasing with temperature $T/n_1 (r) g_1$.
The thermal contribution to the EoS depends also on the system parameters namely $\beta$, $\Delta$ and $\epsilon_j^{dd}$ (see right panel of Fig.\ref{TLHY}). 
The same holds true for the chemical potential of the second component $\delta \mu_2^{th} ({\bf r})$.

The condensed density in Eqs.(\ref{NCDih})-(\ref{EoSih}) can be calculated using the Thomas-Fermi (TF) approximation.
The insertion of corrections (\ref{EoSih}) in the generalized coupled GP equations 
permits us to examine, in a simpler manner, the role of quantum fluctuations in the TF regime. 
Remarkably, the quantum and thermal fluctuations and their corrections to all thermodynamic quantities remain isotropic in the trapped case. 
One can expect  that the inhomogeneity of the system may crucially affect the damping rates and energy shifts of low-energy excitations.
Equation (\ref {EoSih}) will be our starting point in  the next section for analyzing the stability of quantum droplets in dipolar Bose-Bose mixtures 
at both zero and finite temperatures.

\section{Self-bound droplet state} \label{Drop}

Aiming to check the formation of a self-bound droplet state at both zero and finite temperatures, 
we consider a dipolar Bose mixture of intraspecies repulsive interactions, and attractive interspecies interactions. 
The dipole moments of the particles are supposed to be oriented perpendicular to the plane. 
The mutual contact interactions can be tuned via the Feshbach resonances \cite{Roy}.

\subsection{Zero temperature case: Gross-Pitaevskii equation}

At zero temperature, where $\tilde n_j \ll n_{cj}$, $\tilde m_j \ll n_{cj}$, and $N_c \approx N$, the energy density corresponding to the GP equation reads 
\begin{align}\label{Enrgy}
{\cal E}_d &= \sum_j \bigg [\frac{\hbar^2}{2m} |\nabla\Phi_j|^2 +\frac{1}{2} \int d {\bf r' }  V_j ({\bf r-r'})  |\Phi_j({\bf r})|^2  |\Phi_j({\bf r'})|^2 \bigg] \nonumber \\
&+ \int d {\bf r' }  V_{12} ({\bf r-r'}) |\Phi_2({\bf r})|^2  |\Phi_1({\bf r'})|^2 + \delta E_j, 
\end{align}
where $ \delta E_j$ can be calculated from Eq.(\ref{LHYEgy}) or through $\delta E_j=\int \delta \mu_j^0 d n_j$.
Following the procedure outlined in Ref \cite{Petrov}, we assume the two components occupying identical spatial modes $\Phi_j= \sqrt{n_{jc}^{0}} \phi_j $,
with $n_{cj}^{0}$ being the saturation densities. The density ratio which minimizes the energy of the hard mode \cite{Petrov, PetAst, Cab} is given by
\begin{equation} \label{DC}   
\frac{n_{c1}^0}{n_{c2}^0} = \sqrt{ \frac{g_1 (1-\epsilon_1^{dd})}{g_2(1-\epsilon_2^{dd})}}, \;\;\; \;\;\;\ \epsilon_1^{dd}<1\;\; \text{and}  \;\; \epsilon_2^{dd}<1, 
\end{equation}
The condition (\ref{DC}) is necessarily for the stability and the formation of the self-bound droplet. 
In the case of a mixture with tilted dipoles, such a condition becomes dependent on the angle $\theta$.
The dynamics of the dipolar mixture self-bound droplet is described by  the generalized GP equation which can be derived from Eq.(\ref{Enrgy}).
The resulting equation includes an extra LHY repulsive term stabilizing the mixture against collapse.
In the absence of the DDI, it reduces to the Petrov's equation \cite{Petrov}.

From now on, lengths and energies are expressed in units of the extended healing length $\xi$ and $\hbar^2/2m \xi^2$, respectively, where
\begin{equation} \label{HL}   
\xi= \hbar \sqrt{\frac{\sqrt{g_2(1-\epsilon_2^{dd})}/m_1+\sqrt{g_1(1-\epsilon_1^{dd})}/m2} { |\delta {\bar g}| \sqrt{g_1(1-\epsilon_1^{dd})} n_{c1}^0} },
\end{equation}
and $\delta {\bar g}=g_{12}(1-\epsilon_{12}^{dd})+ \sqrt{g_1(1-\epsilon_1^{dd}) g_2(1-\epsilon_2^{dd})}$.
Within these new dimensionless variables, the number of particles is scaled as  $N_c= \xi^3 n_{cj}^{0} {\cal N}$.
For simplicity, we consider the case of a balanced mixture $m_1=m_2=m$, $N_{c1}= N_{c2}=N_c/2$, $a_1=a_2$, and  $\epsilon_1^{dd}=\epsilon_2^{dd}$.
In such a case, the stability and the formation of the droplet state are governed by only the parameters
$a_{12}$,  $\epsilon_{12}^{dd}$, and $N_c$.

We numerically determine the equilibrium state of the dipolar mixture droplet and its energy $E_d=d {\bf r} \int {\cal E}_d (\phi_j , \phi_j^*)$.
Our numerical solution was performed by using a split-step Fourier method 
which has been proven to be a quite powerful numerical tool in solving nonlinear equations (see, e.g.\cite{Pfau, Wach, Boudj77}). 
The DDI terms are treated using a convolution theorem \cite{Pfau1} which allows us to remove the singular nature of the DDI at the origin.
In our case, the LHY term controlled by the functions ${\cal I}_j (\epsilon_{dd}) $ do not require any special adjustment 
since such functions are real for $\epsilon_{ij}^{dd}<1$.
This is in stark contrast with quantum droplets in a strongly dipolar single Bose gas, where LHY quantum corrections need either
a low-momentum cutoff \cite{Saito, Wach} or the lowest-order expansion of the functions $Q_{\ell}(\epsilon_{dd})$ \cite{Pfau, Bess}  in order to avoid 
the imaginary parts.
The algorithm can be checked by reproducing the nondipolar mixture and the single BEC results.

\begin{figure}
\centerline{
\includegraphics[scale=0.85]{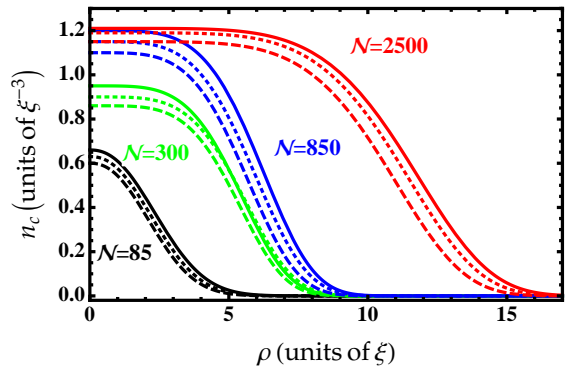}}
 \caption{ (Color online) Density profiles of the dipolar Bose mixture droplet as a function of the radial coordinate for several numbers of particles 
and $a_{12}=-5.5 a_0$ ($a_0$ is the Bohr's radius). 
Solid lines: $\epsilon_{12}^{dd}=0.3$. Dotted lines: $\epsilon_{12}^{dd}=0.5$. Dashed lines: $\epsilon_{12}^{dd}=0.7$.}
\label{ds} 
\end{figure}

Figure \ref{ds} shows that when the number of particles rises, the central density and the radius of the
droplet increase until the system reaches its equilibrium (saturation) in good agreement with the numerical results of \cite{Petrov} 
and with the Monte Carlo simulation predictions \cite{Cik}. 
The equilibrium state  occurs at ${\cal N} \sim 2500$ atoms and $a_{12}=-5.5\,a_0$ indicating that the system is stable 
where only the size dilates and the central density remains constant. 
For ${\cal N} < {\cal N}_{\text{crit}} \approx 85$, the droplet is unstable.
We observe from the same figure that the central density is decreasing with $\epsilon_{12}^{dd}$ which may lead to lowering the critical number of particles.

\begin{figure}
\centerline{
\includegraphics[scale=0.85]{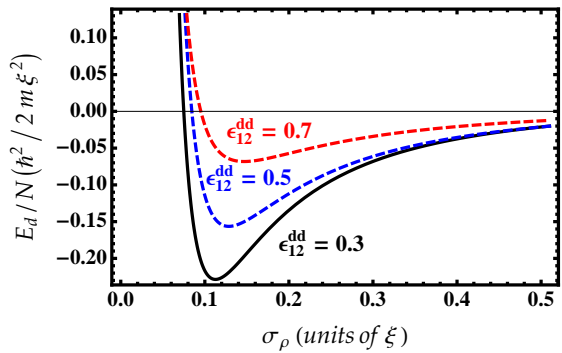}}
 \caption{ (Color online) Energy as a function of the radial size $\rho$ for different values of $\epsilon_{12}^{dd}$.
Parameters are: ${\cal N}=2500$, and $a_{12}=-5.5 a_0$.}
\label{LM} 
\end{figure}

Augmenting $\epsilon_{12}^{dd}$, the strength of the bond is decreasing indicating that
the droplet becomes less stable as is seen in Fig.\ref{LM}.
The change in the energy functional minima persists also in the longitudinal direction (not shown here).
Hence, one can deduce that, for sufficiently large DDI, the local minimum developed in the energy disappears and thus, the mixture droplet undergoes instability.

\begin{figure} 
\centerline{
\includegraphics[scale=0.85]{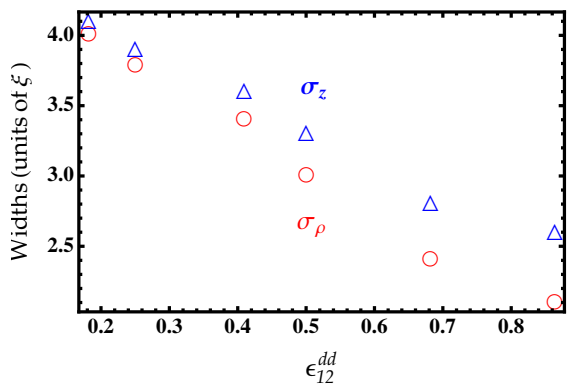}}
 \caption{ (Color online) Longitudinal ($\sigma_z$) and horizontal ($\sigma_{\rho}$) droplet widths as a function of $\epsilon_{12}^{dd}$ for ${\cal N}=2500$.}
\label{Sz} 
\end{figure}

Figure \ref{Sz} depicts that in the equilibrium regime,  the droplet is practically isotropic for small DDI ($\epsilon_{12}^{dd}< 0.1$). 
As $\epsilon_{12}^{dd}$ increases the longitudinal ($\sigma_z$) and horizontal ($\sigma_{\rho}$) droplet widths decrease
and  the droplet is going to be anisotropic owing to the anisotropy of the DDI.
Such an anisotropy becomes important for large $\epsilon_{12}^{dd}$, for instance for $\epsilon_{12}^{dd}=0.8$, $\sigma_z/\sigma_{\rho} \simeq 1.25$.
The widths can be extracted from the extended GP equation solutions employing $\sigma_i^2= c_i\int r_i ^2 |\phi ({\bf r})|^2 d {\bf r}$, where $i=z, \rho$, 
and $c_i$ are some normalization constants.  

\subsection{Finite temperature: TDHFB equations}

Now we extend our study for a mixture droplet with DDI to finite-temperature using our full TDHFB equations
which include  in addition to the standard LHY term (\ref{EoSih}) another extra term related to the thermal LHY corrections (\ref{EoSihT}). 
Our objective is to look at how the equilibrium is affected by such thermal fluctuations.  
The energy functional (\ref{Enrgy}) must acquire temperature dependence.
Here we recall that the long-range exchange terms  $\tilde n ({\bf r},{\bf r'})$ and $\tilde m ({\bf r},{\bf r'})$ are neglected 
since they are not important for ${\bf r} \neq {\bf r'}$ \cite {Bon, BoudjDp}.

\begin{figure}
\centerline{
\includegraphics[scale=0.85]{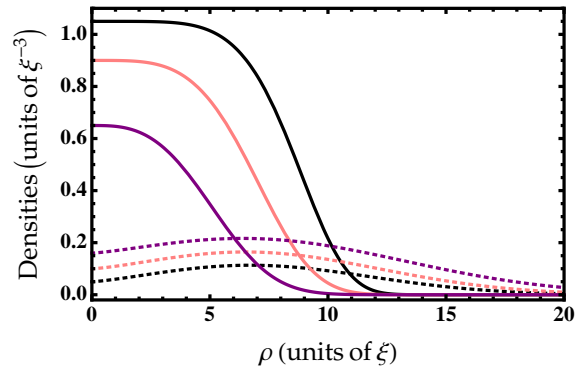}}
 \caption{ The condensed (solid) and noncondensed (dotted) densities versus the radial distance for several values of temperatures.
Parameters are: ${\cal N}=2500$, $\epsilon_{12}^{dd}=0.5$, and $a_{12}=-5.5 a_0$.
Black lines: $T=50$ nK. Pink lines: $T=90$ nK. Purple lines: $T=150$ nK. 
The noncondensed density has been amplified ten times for clarity.}
\label{dsT} 
\end{figure}

Figure.\ref{dsT} depicts the condensate and noncondensate density profiles in the droplet for
a range of temperatures below the transition temperature 
(the temperature at which the total number of atoms becomes comparable to the number of noncondensed atoms). 
We see that the noncondensed density increases with increasing temperature whereas the condensed density is reduced due to the thermal fluctuations. 
As the temperature rises, the atoms evaporate out of the self-bound droplets due to dissipation forming a broader thermal halo  (a peak) near the edge of the condensate
that results in the saturation density, and the critical number of particles are lowered which may cause destabilization of the droplet.
We note that a similar behavior holds in a single-component dipolar droplet \cite{BoudjDp}.

\begin{figure}
\centerline{
\includegraphics[scale=0.85]{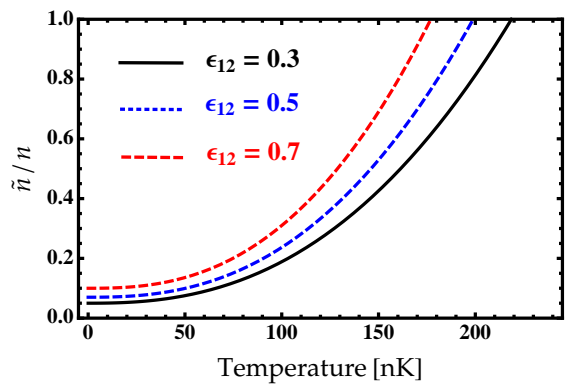}}
 \caption{ Condensed depletion $\tilde n/n$ as a function of $T/n_1g_1$ for several values of $\epsilon_{12}^{dd}$.
Parameters are the same as those in Fig.\ref{LM}. }
\label{dp} 
\end{figure}

In Fig.\ref{dp}, we show the temperature dependence of the condensed depletion $\tilde n/n$.
At zero temperature, $\tilde n/n$ is $\sim 5 \%$ for $\epsilon_{12}^{dd}=0.3$ and does not exceed
$10 \%$ for the parameters considered above.  
For $T< \,50$ nK,  the depletion  depends weakly on temperature, while the situation is inverted at higher temperatures. 
At fixed temperature, the condensed depletion is growing with DDI, for instance, at $T \simeq 150$ nK,  it augments by $\sim 20\%$ 
when  $\epsilon_{12}^{dd}$ varies from 0.3 to 0.7.

\section{Conclusion and outlook} \label{concl}

We studied the properties of two-component dipolar Bose condensates at nonzero temperatures. 
We showed that such a system features remarkable properties. 
Coupled equations of motion have been derived to describe, in a self-consistent way, the dynamics of the condensates. 
These equations can be considered as a finite-temperature extension of the standard coupled GP equation for dual condensates with DDIs.
In the case of the homogeneous mixture, the shift to the excitations, the chemical potential, the ground-state energy, and 
the compressibility due to quantum and thermal fluctuations corrections has been precisely determined. 

We showed that the developed method is a powerful tool for investigating the quantum droplet state
in a dipolar bosonic mixture with intraspecies repulsive interactions and attractive interspecies interaction.
We pointed out that, when the contact interaction dominates the dipolar one, the droplet is stabilized due to 
the repulsive first-order LHY corrections.  The anisotropy  of the DDI shapes the droplet anisotropic geometry.
The properties of mixture quantum droplets such as the stability, the density profiles, the energy,  and the widths
have been found to be modified owing to the intriguing role of the DDIs. 
We presented also a detailed analysis of the temperature dependence of the condensate and noncondensate density profiles of the self-bound droplet.
Our results revealed  that as the temperature and the DDI are increased, atoms leave the droplet forming a thermal cloud surrounding the condensate. 
This process continues without ceasing until the temperature reaches its critical value above which the droplet becomes unstable.

We hope that the findings of this paper will be useful for inspiring future experiments on mixture droplets with DDIs. 
An interesting future application of our TDHFB theory includes the study of the effects of  both DDI and temperature on the collective modes of a mixture droplet, 
and checking whether the self-evaporation of such a state predicted in Ref \cite{Petrov} still remains.
Another important aspect is to investigate the formation of a droplet state in a mixture with 
repulsive intraspecies and attractive interspecies interaction and strong DDI ($\epsilon_i^{dd}>1$ and $\epsilon_{ij}^{dd}>1$). 
It is an open question whether quantum fluctuations can arrest collapse originating from both attractive forces and dipolar interactions.

\section{Acknowledgements}

We thank Dmitry Petrov for insightful discussions.

\end{document}